\documentclass[twocolumn,twoside,10pt]{article}
\usepackage[a4paper,top=2.5cm,bottom=2.5cm,left=2.5cm,right=2.5cm]{geometry}
\usepackage{times}
\usepackage{natbib}
\usepackage{amssymb}
\usepackage{amsmath}
\usepackage{amsfonts}
\usepackage{epsfig}
\usepackage{graphicx}
\usepackage{color}
\usepackage[font=small]{caption}

\usepackage{epstopdf}

\newcommand{\be}{\begin{equation}}
\newcommand{\ee}{\end{equation}}
\newcommand{\ba}{\begin{eqnarray}}
\newcommand{\ea}{\end{eqnarray}}
\newcommand{\bd}{\begin{displaymath}}
\newcommand{\ed}{\end{displaymath}}

\begin{document}

\title{Volume Ignition via Time-like Detonation in Pellet Fusion}

\author{L.P. CSERNAI$^1$ 
\thanks{Email: csernai@ift.uib.no},
\textsc{and}
D.D. STROTTMAN$^2$\\
\small{$^1$ Institute of Physics and Technology, University of Bergen,
Allegaten 55, 5007 Bergen, Norway}\\
\small{$^2$ Los Alamos National Laboratory, Los Alamos, 87545 New Mexico, USA}\\%
}
\smallskip
\date{\today}
\maketitle
%
{\bf Abstract:}
Relativistic fluid dynamics and the theory of relativistic detonation
fronts are used to estimate the space-time dynamics of the burning
of the D-T fuel in Laser driven pellet fusion experiments. The 
initial ``{\it High foot}" heating of the fuel makes the compressed
target transparent to radiation, and then a rapid ignition pulse can
penetrate and heat up the whole target to supercritical temperatures
in a short time, so that most of the interior of the  target ignites 
almost simultaneously and instabilities will have no time to develop.
In these relativistic, radiation dominated processes both the interior,
time-like burning front and the surrounding space-like part of the front
will be stable against Rayleigh-Taylor instabilities. To achieve 
this rapid, volume ignition the pulse heating up the target to
supercritical temperature should provide the required energy in less
than ~ 10 ps.   
%
\smallskip


\bigskip

Recent experimental and theoretical efforts to achieve Inertial Confinement
Fusion (ICF) have focussed on the compression of the fuel in a 2.263 mm 
diameter 
fuel capsule including a CH ablator, a thin D-T ice layer, and a D-T gas 
fill in the middle 
\citep{NIF14}; \citep{NIF14prl}. 
The ablator layer was placed on the 
external surface to achieve a larger compression of the pellet. The outcome of 
these experiments was that most of the pellet broke into pieces due to 
Rayleigh-Taylor surface instabilities, which are common in surfaces
under extreme dynamical pressure.

Already in the classical literature, 
\citep{ZR1969} mentioned
that the way to avoid these instabilities is to make the compression and the 
detonation front a high temperature radiation dominated front, which works to
smooth out the Rayleigh-Taylor (RT) instabilities propagating with the 
sound-speed, while the radiation is propagating with the speed of light.

From this well known reason it must be obvious that the radiation dominated,
high temperature process must be described with relativistic fluid dynamics,
wherein the pressure is not neglected compared with the energy density, and the 
propagation of radiative energy is described in a consistent way with
all other dynamical processes. 

Unfortunately the overwhelming majority of theoretical models
that attempt to describe ICF experiments are (i) 
non-relativistic, i.e. neglecting pressure, and (ii) radiation is assumed to 
have infinite speed in principle. (Of course such an ``infinite speed" becomes
an undefined speed, which depends on the space-time grid resolution and on the
numerical method.)

In the present work we concentrate on the ``volume ignition"of the fuel by
neglecting (for the 1st approximation) compression. We use a relativistic 
Rankine-Hugoniot description originally described by 
\citep{Taub48}, 
which description was then 39 years later corrected by  
\citep{Cs87}, and used since then
widely in the field of relativistic heavy ion collisions 
\citep{Cs94}.
Just as in the Rankine-Hugoniot non-relativistic description of shock 
waves the relativistic
relations are also based on the energy-momentum tensor, $T^{\mu\nu}$, 
and baryon charge current, $N^{\mu}$, 
conservation across a hyper-surface with a normal 4-vector $\Lambda^\nu$,
where the change of a quantity $a$ across the hyper-surface is denoted by
$[a]=a_2 - a_1$:

\be
[R^\mu]=[T^{\mu\nu} \Lambda_\nu] = 0,\ \ \ \ \ \
[j]=[N^{\mu} \Lambda_\mu] = 0.
\label{e1}
\ee
These conservation laws lead to the relativistic shock or detonation 
equations for the energy density, $e$, pressure, $p$, and generalized 
specific volume, $X= (e+p)/n^2$:
\be
         j^2 =  \Lambda^\mu \Lambda_\mu \, [p] / [X], 
\nonumber
\ee
\be
       [p] (X_1{+}X_2) = [(e+p) X].
\label{e2}
\ee
This description treats detonations also across hyper-surfaces with
time-like normal vectors ($\Lambda^\mu \Lambda_\mu = {+}1$),
and therefore has the name time-like detonation,
which actually means simultaneous volume ignition. 

E.g. this description
gives a correct, and rigorous description of the sudden and rapid 
hadronization of the Quark-Gluon Plasma (QGP) with the release of large latent
heat of the QGP to hadronic matter phase transition 
\citep{Cs87}, \citep{Cs94}.
Taub's description could 
be applied to ``slow", space-like fronts only.

This simplified model gives a quantitative estimate of the dynamics of
volume ignition, which occurs in such a time-like detonation, thus completely
avoiding the possibility of RT type of mechanical instability.

In the work of \citep{Cs87}, the compression of the fuel in a sphere
is neglected and the heating is described by isotropic radiation inwards from
all sides, where $Q$ is the heat radiated inwards in unit time per unit 
surface. The opacity of the target is assumed to be uniform inside the sphere,
a fraction, $C$, of the radiated heat is absorbed in the fuel 
and it is assumed that the incoming energy is sufficient that it is absorbed
by the fuel by the point when the incoming beam reached the surface of the 
fuel on the back side of the pellet and not before. Under these assumptions 
the temperature increases and reaches a critical (ignition) temperature
at some space-time point, $T(r,t) = T_c = 4 \pi Q C / C_v$, where $C_v$ is
the specific heat.

\begin{figure}  
\begin{center}
\resizebox{0.9\columnwidth}{!}
{\includegraphics{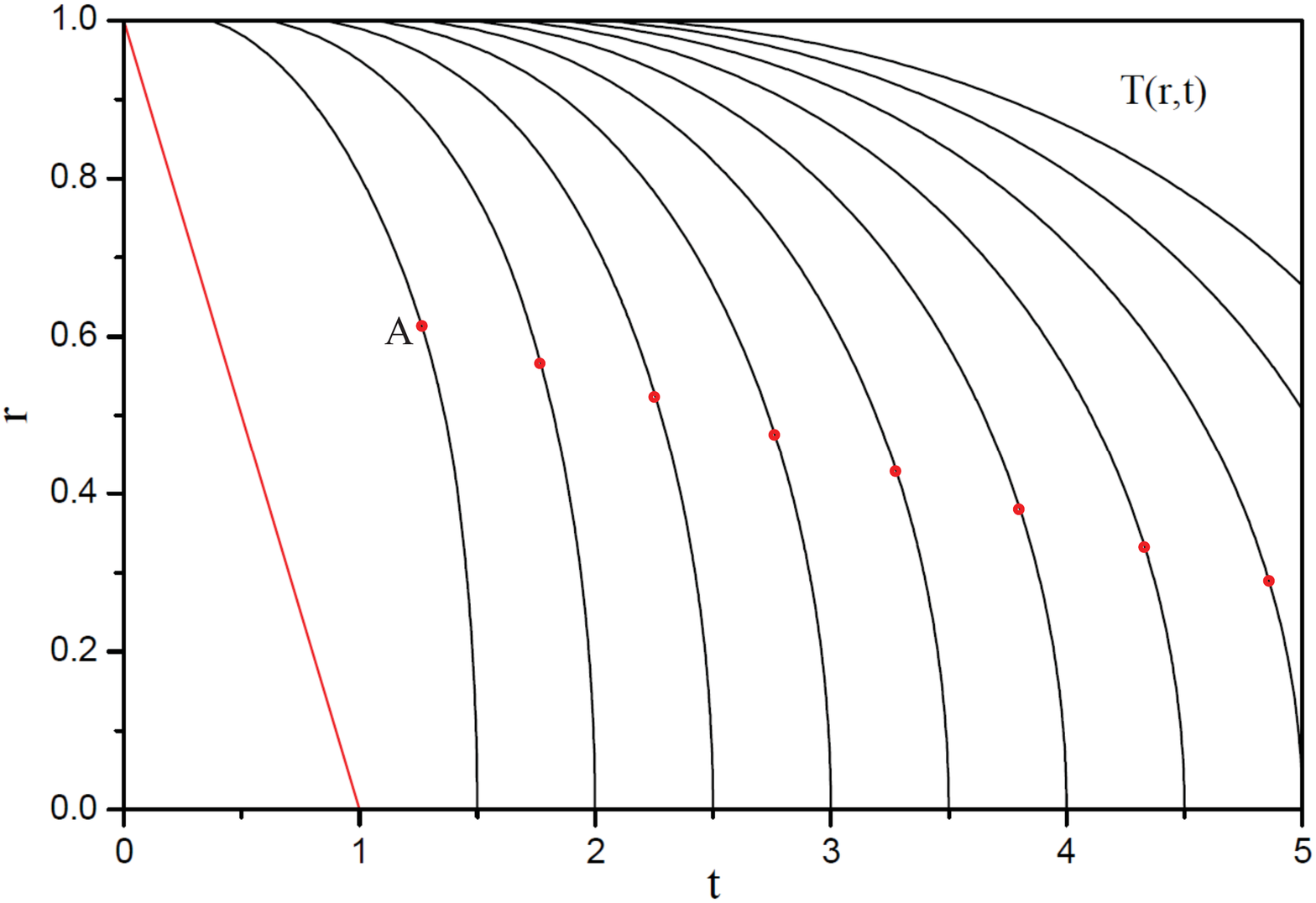}}
\caption{
(color online)
Acceleration of the detonation fronts where the temperature, $T$, 
reaches $T_c$ due to radiation from the outer 
surface at $r=1$ inwards (to $r=0$). The heating by radiation can lead 
to a smooth transition from space-like to time-like front at the red points
A, where the front propagates with the speed of light. This is possible 
because the fluid does not move together with the front. 
The section of the detonation front outside the point A ($r>r_A$) 
is a space-like front, while the section within A ($r<r_A$)
 is time-like, which corresponds to
volume ignition.  The ten contour lines indicate the space-time points 
where $T(r,t) = 1,\ 2,\ ...\ 10 T_c$. If the radiation heating is so fast
that ignition is reached at  $T(r,t) = T_c$ then the internal domain
of the front, up to $r \approx 0.6$, is time-like or in this domain rapid
volume ignition takes place. To achieve this the heat needed for ignition
should be absorbed by the fuel within $t_{in} = 1$. If the irradiation
is slower, so that ignition is reached only at $T(r,t) = 10\, T_c$ or later, 
then the domain of volume ignition is negligible, and the dominant space-like 
detonation is subject to mechanical RT instabilities.
The world lines of the fluid particles are horizontal straight lines.
The inward light cone is indicated with the red straight
line starting from
$t=0,\ r=1$ and reaching the center at $t=1$.  Here 
$t$ is measured in units of $R_0/c$,  $r$ in units of
$R_0$, and $T$ in units of $2\pi C Q/C_V$. 
Based on the work of \citep{Cs87}.
}
\label{F_1}
\end{center}
\end{figure}

Under these assumptions an analytic solution is given 
by \citep{Cs87}, providing the
space-time points of a constant temperature hyper-surface, see FIG. \ref{F_1}.  
\be
T(r,t) \propto \left\{
\begin{array}{ll}
0 \ ,\ \ \ \ \ \ \ \ t{<}1{-}r \\
\frac{t}{r} \left( \ln \frac{t}{1-r} - 1 \right) + \frac{1-r}{r}\ , \\
 \phantom{mmmmmm} 1{-}r{<}t{<}1{+}r  \\
\frac{t}{r} \ln \frac{1+r}{1-r} - 2 \ ,\ \ \ \ \ \ t{>}1{+}r
\end{array}
\right.
\ee

The isotherm hyper-surface is such that in an external spherical layer the
detonation starts from the outside with sub-luminous velocity propagating 
inwards. This ``propagation" velocity increases, reaches $c$,
the velocity of light, at a radius $R_A$
(indicated by red dots on the $T=$ const. contour lines, FIG. \ref{F_1}).

Then the points of the detonation hyper-surface inside this radius are not
in causal connection with each other and the ignition completes with a 
semi-simultaneous volume ignition. This process is not acausal, it is the
consequence of the initial inward radiation, so that all points of the
time-like detonation hyper-surface are within the light-cones of the energy
emission at the initial external surface sphere of radios $R_0$.
The idea of volume ignition was presented earlier based on different
arguments by 
\citep{KCHS89}, with a short laser ignition pulse.

This model became relevant and applicable to the recently published ICF
experiments performed at the National Ignition Facility (NIF). 
To achieve a rapid volume ignition the needed total ignition energy should be
radiated inward in a time interval, $t_{in} < 1$ or  $t_{in} \ll 1$ 
(in units of $[R_0/c]$).

Of course the model is oversimplified. Even if we irradiate the fuel with 
light the radiation pressure is one third of the radiation energy density, 
$p_R = e_R/3$, and this presses the fuel before ignition. But this lasts for a
short enough time to avoid the build-up of a RT instability. The mechanical RT
instability propagates with a speed much smaller than the speed of light,
$ 10^4 - 10^5$ times less in a solid, and about 10 times smaller at nuclear 
matter density.

The longer is $t_{in}$ the greater probability we have for RT instability. If 
$t_{in} > 3 $, the RT instability can hardly be avoided, and the possible 
volume ignition domain size becomes negligible.
Thus for an $R_0 = 3$ mm pellet the ideal irradiation time for volume ignition
would be $t_{in} \le 10$ ps  (while for an $R_0 = 30$ cm target it would be
$t_{in} = 1$ ns).

At the NIF ``High foot" pre-compression of   
the fuel capsule in fortunate configuration will not lead to 
RT instability. As presented by \citep{NIF14} this shorter, $\sim 6$ ns, 
``High foot" pre-compression led to a more isotropic polar view 
(Figure 2b in ref. \citep{NIF14}) giving rise to less RT instability
as it is shown by
\citep{Casey14}, and higher total yield.
As also mentioned in ref. \citep{Atzeni14}, 
the pre-compression stage
should be concluded by an ignition spike of 10-30 ps. This is not very far
from our suggestion of 10 ps or less to achieve volume ignition
or a time-like detonation. In ref. \citep{Atzeni14} 
this is not considered
and only conventional shocks are discussed in the form of an imploding shock
followed by a rebounding outgoing shock. These mechanical, pressure
driven processes are still subject to RT instability, while the somewhat
shorter and more energetic irradiation can prevent the possibility
of all mechanical instabilities.

In the present ignition experiments a compression
stage precedes the ignition process 
\citep{NIF14}, \citep{NIF14prl}.
In the most successful experiments with an initial ``High foot" shock
of $\sim 6$ ns the capsule is heated up and compressed to a smaller
size of a radius of about an order of magnitude smaller, making the 
target hotter and more dense. The hotter target became less opaque,
and the target more compressed. Then two final, more energetic shocks
lasting for $\sim 8$ ns, resulted in a hot spot of  $\sim 37\ \mu$m
radius, which is (partly) ignited and emitted almost the same energy
that was invested to its compression and heating. The rapid ignition of an
$R_0 = 300\ (100)\ \mu$m target to ignition at a mostly time-like surface
would need a short energetic pulse of $t_{in} \approx 1 (0.3)$ ps, which
is about $10^3 - 10^4$ times less than it was done in the NIF
experiment.  

\begin{figure}  
\begin{center}
\resizebox{0.9\columnwidth}{!}
{\includegraphics{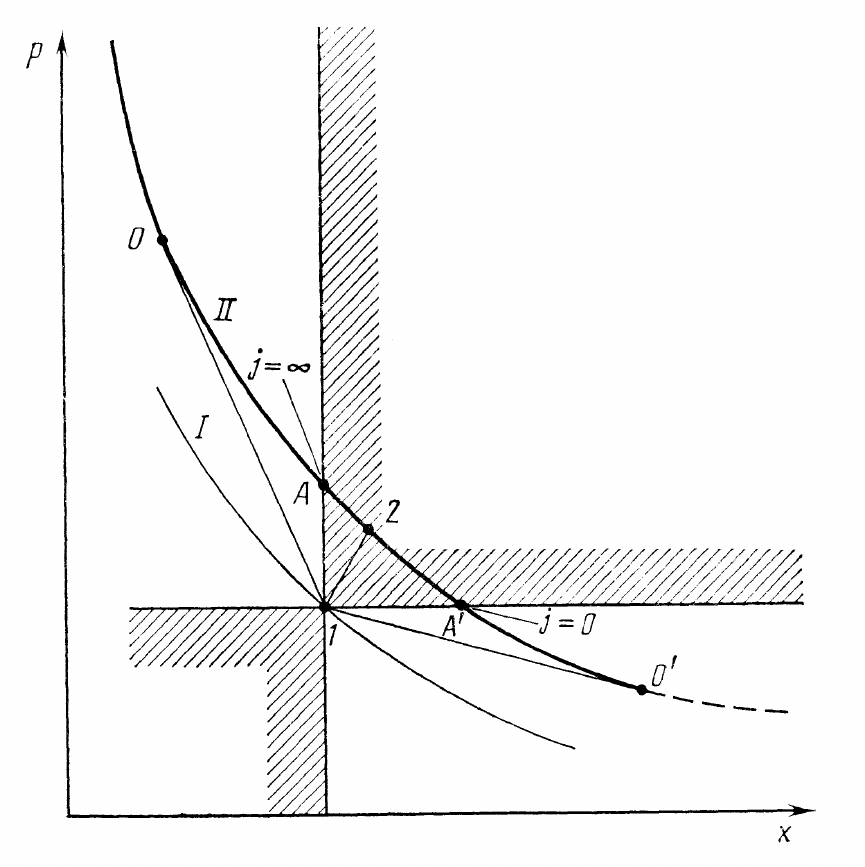}}
\caption{
The shock and detonation adiabats in the pressure, p, generalized specific
volume, X, plane, where the initial state is labeled by $1$. The shock adiabat
(I) goes through point $1$, the detonation adiabat (II) represents the
final state after the detonation with another Equation of State, which 
already contains the latent heat of the detonation process,
and therefore it is at higher p and X than the shock adiabat. The lines
starting from point $1$ are the Rayleigh lines connecting
the initial and final states. Their slope is determined by the current across
the front, $j$. Mechanically stable final states are above the 
Chapman-Jouget point, $O$, while the section of the adiabat between points
$O$ and $A$ are mechanically unstable, and the points between $A$ and $A'$
were considered unphysical. Radiation dominated processes may stabilize
final states in the section between $O$ and $A$, and enable the existence
of final states between $A$ and $A'$ in time-like detonations or in
simultaneous volume ignition.
From the work of \citep{Cs87}. 
For a more detailed discussion see the book by \citep{Cs94}.
}
\label{F_2}
\end{center}
\end{figure}

Nevertheless the radiation dominated implosion leads to increased stability
in the space-like, mechanically unstable section of the detonation 
adiabat under the Chapman-Jouget point (point ``O" in Figure \ref{F_2}, and
see also 
the discussion in Chapt. 5 of the textbook by \citep{Cs94} also). Thus, the 
shortening of the pulse of the last two shocks, without 
decreasing the total deposited energy, leads to radiation
dominance and may increase the stability considerably 
compared to the relatively ``slow" final compression and
heating of $\sim 8$ ns.

Apart from the (i) short and energetic ignition pulse, the other critical 
condition is the (ii) sufficiently small opacity, such that the ignition pulse 
should be able to penetrate the whole target. Luckily the more energetic
preheating pulses lead to a smaller target opacity and better 
thermal conductivity 
as shown by \citep{sxHu14},
so the appropriate tuning of the laser irradiation must be possible. The 
present pre-compression and heating with the starting ``High foot" pulse
led to higher temperature after the pre-compression, and this may be 
sufficient or adequate to achieve the optimal (low) opacity for the final
short ignition pulse.  The presented simple analytic model can be best
applied to the stage of the process after the  ``High foot" pulse heated
up the fluid sufficiently to achieve sufficiently small opacity.

Both these conditions may be more easily satisfied by heavy ion beams, where
the beam bunch length can be regulated by beam optics and could be adjusted so 
that the length at the target becomes as short as $10$ ps, (or $3$ mm) or less.
The suggestions of \citep{H14} and \citep{Hea14c} emphasize
also fast ignition as this work, however, this is considered
to happen in a well focussed location with subsequent spreading the
fusion flame through the whole (pre-compressed) target. The 
expansion of the target is counteracted by a strong magnetic field
generated by a previous laser pulse. The primary aim is to generate
as strong mechanical compression as possible. This is not the 
mechanism suggested here. 

The detailed study of \citep{Fernandez} has basically the same goal,
although instead of the magnetic field the possibilities of using
proton and ion beams are discussed. It is even discussed that 
to reach maximal acceleration and relativistic speeds in the implosion,
the plasma should be made opaque to the extent that it 
``becomes sufficiently thick to isolate the rear target surface from the
laser".  

%

The problem with these approaches is that this extreme
compression leads to RT instabilities, which 
leads to a strong rebound of most of the target material. (In a 
collapse of a cold, $\sim$ 20 Solar mass Supernova progenitor only up to
a 2 Solar mass Neutron or Hybrid star remains compressed.) 
To describe these instabilities a very high resolution, 3+1 dimensional
relativistic PICR code is necessary, \citep{CSA12} with very small
numerical viscosity. Using such a relativistic approach one can see that
the explosive hadronization of QGP takes place in a large part of the
QGP volume with an almost simultaneous detonation, see the works of
\citep{FW14} and \citep{CC09}.  This is essentially the only secure
way to prevent instabilities, and this requires a radiation dominated
relativistic process, as shown also in Figure \ref{F_1}. To achieve this
almost {\em simultaneous} ignition in a large part of the
pre-heated target of $\sim 100 \mu$m, this target should not be further
compressed, rather heated up by radiation, possibly simultaneously from
all sides, so that radiation can reach also ``the rear target surface".
Thus, in this aspect, the present suggestion is an alternative of the
focussed single point fast ignition discussed by
\citep{Fernandez}, \citep{H14} and \citep{Hea14c}.
On different grounds it was also found that picosecond pulses may
lead to volume ignition.
This in any case 
excludes the development of most instabilities 
as shown by \citep{HMLM}, \citep{HKM14} and \citep{H13}.

The penetration length of heavy ion beams can also be regulated by the 
well-chosen beam energy. This operation is well studied in connection with 
radiation therapy where the position of the absorption Bragg peak is tuned to
the necessary depth in the body to reach the tumor. One could also study if
in this case one can achieve a sufficiently short and energetic final pulse 
so that the detonation becomes time-like (in the section of the detonation
adiabat between points A and A' in FIG. \ref{F_2}) in 
the majority of the compressed target,

Enlightening discussions with Dujuan Wang, Dieter H.H. Hoffmann and
Heinrich Hora are gratefully acknowledged.


\end{document}